\documentclass[12pt,a4paper]{article}
\usepackage[dvipdfmx,hiresbb]{graphicx}
\usepackage[dvipdfmx]{color}

\usepackage{color}
\usepackage{bm,enumerate,amsmath,amssymb,amsthm}
\usepackage{epsfig}
\usepackage{cite}
\usepackage{algorithm}
\usepackage{algorithmic}
\usepackage{txfonts}
\usepackage{subfigmat}
\usepackage{figsize}
\usepackage{here}
\usepackage{listings}
\usepackage{braket}
\usepackage{comment}
\usepackage{lscape}
\usepackage{bigints}

\usepackage{arxiv}
\usepackage[utf8]{inputenc} % allow utf-8 input

\theoremstyle{definition}

\newtheorem*{theorem*}{Theorem}

\newtheorem*{definition*}{Definition}

\renewcommand{\headeright}{}

\title{Approximate Cosine Similarity Estimation based on an Angle-Encoding Hadamard Test}

\author{
  Hiroshi Ohno \\
  Toyota Central R \& D Labs., Inc.\\
  Aichi, Japan \\
  \texttt{oono-h@mosk.tytlabs.co.jp} \\
}

\date{\empty}

\begin{document}

\maketitle

\begin{abstract}
  The Hadamard test is a standard quantum primitive for estimating inner products and expectation values, but in data-processing settings its practical utility is often limited by the cost of preparing amplitude-encoded quantum states.
  In this study, we investigate an angle-encoding variant of the Hadamard test for estimating cosine similarity between normalized real-valued vectors.
  The proposed method decomposes the similarity computation into elementwise two-qubit Hadamard-test circuits that can, in principle, be executed in parallel, resulting in constant circuit depth with respect to the vector dimension at the expense of a larger qubit footprint and classical post-processing.
  Because the resulting estimator is approximate, we analyze the induced bias and show that it is non-negative under the approximation used in our derivation.
  Numerical experiments on random normalized vectors show that, in the tested setting, the estimation error decreases as the vector dimension increases.
  We further illustrate a possible application to cosine-attention-based Transformer models.
  These results suggest that the angle-encoding Hadamard test may provide a useful design point for near-term similarity estimation when shallow circuit depth is preferred over compact qubit usage.
\end{abstract}

\section{Introduction}
The Hadamard test is one of the most fundamental subroutines in quantum computing for estimating the real or imaginary part of matrix elements of the form $ \braket{\psi | U | \psi} $.
In quantum machine learning and quantum-enhanced data processing, this primitive is frequently used to evaluate similarities, overlaps, kernels, and related quantities derived from quantum state overlaps.
However, when classical data must first be embedded into quantum states, the overall advantage of the Hadamard test can be substantially reduced by the cost of state preparation.

A common strategy is to encode a $ d $-dimensional classical vector into amplitudes of an $ \mathcal{O}(\log d)$-qubit state.
For example, the M\"{o}tt\"{o}nen algorithm \cite{mottonen2005} is often used for state preparation.
While this representation is compact in qubit count, exact amplitude encoding of unstructured classical data can require expensive state-preparation circuits, making the full workflow less attractive on near-term quantum (NISQ) devices.
In particular, standard state-preparation methods based on uniformly controlled rotations illustrate that low qubit count does not automatically imply low implementation cost.
For this reason, it is worthwhile to investigate alternative encodings that trade qubit efficiency for shallower circuits and simpler data loading.

In this study, we investigate an angle-encoding variant of the Hadamard test for estimating cosine similarity between normalized real-valued vectors with entries in $ [-1,1] $.
Instead of preparing an amplitude-encoded global state, we encode each vector entry locally through a single-qubit rotation and evaluate the contribution of each component using an elementwise Hadamard-test circuit.
This leads to a parallel construction with constant circuit depth with respect to the vector dimension, provided that sufficient qubits are available.
The cost for this simplification is twofold: the method uses more qubits than amplitude encoding, and the resulting cosine-similarity estimator is approximate rather than exact.

Our goal is not to claim a universal replacement for conventional amplitude-encoding-based Hadamard tests.
Rather, we identify a practically relevant operating point in which shallow depth and simple local data loading are prioritized.
Within this setting, we derive the approximate estimator, discuss its bias, and numerically evaluate its behavior on random normalized vectors.
We also show that the approximation used in the estimator induces a non-negative bias.
Additionally, as a demonstration application, we examine how the proposed estimator can be incorporated into cosine-attention-style Transformer computations.

The main contributions of this study are as follows.
First, we formulate an angle-encoding, elementwise-parallel Hadamard-test construction for approximate cosine-similarity estimation.
Second, we analyze the bias introduced by the approximation used to eliminate the non-polynomial square-root term.
Third, we provide numerical evidence that the approximation error decreases in the tested regime as the vector dimension increases.
Finally, we demonstrate a preliminary application to a Transformer model employing cosine attention (cottention).

We emphasize that the present method is intended as an approximate, shallow-depth similarity estimator under bounded real-valued inputs, rather than as a general exact substitute for amplitude-encoding-based Hadamard tests.

The remainder of this paper is organized as follows.
In Section \ref{sec2}, we describe the relationship to the most relevant recent work \cite{mehta2025}.
Section \ref{sec3} introduces the proposed method and its bias analysis.
Section \ref{sec4} reports numerical results, including both random-vector experiments and a demonstration Transformer application.
Section \ref{sec5} summarizes this study.

\section{Related work}\label{sec2}
Several variants of the quantum Hadamard test have been proposed to address the practical limitations of amplitude encoding in quantum machine learning.
In particular, Mehta et al. \cite{mehta2025} introduced the Generalized Quantum Hadamard Test (GQHT), which extends the conventional Hadamard test to compute inner products between bounded real-valued vectors without requiring L2 normalization.
Their approach relies on a nonlinear quantum feature mapping involving additional component and utility qubits, and enables exact inner-product estimation within the bounded input space.
GQHT has been shown to be effective as a subroutine in various hybrid quantum-classical machine learning models, including logistic regression and centroid-based classifiers.
In contrast to these works that primarily focus on generalizing the class of computable similarity measures while preserving exactness, our study explores a different design objective.
We investigate an angle-encoding Hadamard test that prioritizes constant circuit depth and simple local data loading over exact estimation.
By decomposing the similarity computation into elementwise, parallel Hadamard-test circuits, our method avoids global amplitude encoding at the cost of introducing a controllable approximation bias.
This positions our method as a complementary alternative to GQHT constructions, targeting NISQ-era settings in which shallow circuit depth and reduced coherence requirements are of primary importance.

Our contribution differs from prior works on GQHT constructions in the following aspects:
\begin{itemize}
\item We propose an angle-encoding, elementwise-parallel Hadamard-test construction that achieves $ \mathcal{O}(1) $ circuit depth with respect to vector dimension, rather than focusing on exact inner-product estimation.
\item Our method explicitly trades estimation accuracy for circuit shallowness and locality, resulting in an approximate cosine-similarity estimator with a provably non-negative bias.
\item Unlike GQHT schemes that employ global nonlinear feature mappings, our method relies solely on local single-qubit rotations and two-qubit Hadamard-test primitives, making it particularly suitable for NISQ devices.
\item We demonstrate the practical implications of this design choice through numerical experiments and a demonstration application to cosine-attention-based Transformer models.
\end{itemize}

In summary, while GQHT approaches aim to extend the class of similarity measures that can be computed exactly, our study investigates the opposite regime, in which circuit depth and implementation simplicity are prioritized over exactness, leading to an approximate but highly parallelizable cosine-similarity estimator tailored to NISQ-era constraints.

\section{Method}\label{sec3}
\subsection{Conventional Hadamard test}
When a controlled-unitary $ U $ is given, the Hadamard test produces the following result:
\begin{equation}\label{eq1}
  {\rm Re} \braket{0 | U | 0},
\end{equation}
where the initial state is $ \ket{0} $.
To realize angle encoding using the Pauli Y-rotation gate $ R_{y} (\theta) $, $ U $ is constructed by
\begin{equation}\label{eq2}
  U = R_{y}^{\dagger}(\theta_{x}) R_{y}(\theta_{y}),
\end{equation}
where $ \theta_{x} $ and $ \theta_{y} $ denote the angle corresponding to the real-valued data $ x $ and $ y $, respectively.
In particular, $ \theta_{x} = 2 \arccos(x) $ ($ -1 \leq x \leq 1 $).
Additionally, $ U $ is converted into a controlled-unitary, which depends on the state of an ancilla qubit.

\subsection{Proposed algorithm}
In order to calculate the inner product (cosine similarity) between two vectors, we present an angle-encoding parallel elementwise Hadamard test.
Note that for two vectors $ v $ and $ w $,
\begin{equation}
  Similarity(v, w) = \cos(\theta) = \frac{v \cdot w}{\| v \| \, \| w \|},
\end{equation}
where $ \theta $ is the angle between $ v $ and $ w $.
For the two $ i $-th elements $ v_{i} $ and $ w_{i} $ of two real-valued normalized vectors, the outcome of the Hadamard test is given by
\begin{equation}\label{eq3}
  {\rm Re} \braket{0 | U_{i} | 0} = (v_{i} \bra{0} + \sqrt{1 - v_{i}^{2}} \bra{1})(w_{i} \ket{0} + \sqrt{1 - w_{i}^{2}} \ket{1}).
\end{equation}
Then, we have
\begin{equation}\label{eq4}
  v_{i} w_{i} = {\rm Re} \braket{0 | U_{i} | 0} - \sqrt{1 - v_{i}^{2}} \sqrt{1 - w_{i}^{2}}.
\end{equation}
Moreover, for the second term on the right-hand side, when $ v_{i}^{2} $ and $ w_{i}^{2} $ are small, a first-order approximation gives
\begin{equation}\label{eq5}
  \begin{split}
    \sqrt{1 - v_{i}^{2}} \sqrt{1 - w_{i}^{2}} &\approx 1 - \frac{v_{i}^{2} + w_{i}^{2}}{2} + \frac{v_{i}^{2} w_{i}^{2}}{4} - \frac{v_{i}^{4} + w_{i}^{4}}{8}\\
    &\approx 1 - \frac{v_{i}^{2} + w_{i}^{2}}{2}.
  \end{split}
\end{equation}
Therefore, the cosine similarity between $ v $ and $ w $ is given by
\begin{equation}\label{eq6}
  \begin{split}
    Similarity(v, w) &= \sum_{i=1}^{d} v_{i} w_{i} = \sum_{i}^{d}{\rm Re} \braket{0 | U_{i} | 0} - 1 + \frac{v_{i}^{2} + w_{i}^{2}}{2}\\
    &= \left( \sum_{i=1}^{d} {\rm Re} \braket{0 | U_{i} | 0} \right) - d + 1,\\
  \end{split}
\end{equation}
where we used $ \| v \| = \| w \| = 1 $ and $ d $ denotes the vector dimension.

$ {\rm Re} \braket{0 | U_{i} | 0} $ is realized by a two-qubit quantum circuit in which a two-qubit controlled unitary $ U_{i} $ is used.
Therefore, for a $ d $-dimensional vector, $ 2d $ qubits are needed.
However, the circuit depth is $ \mathcal{O}(1) $, and the implementation cost can be significantly reduced compared with the conventional Hadamard test based on amplitude encoding.

\begin{figure}[htbp]
  \centering
  \begin{tabular}{c}
    \includegraphics[width=8cm]{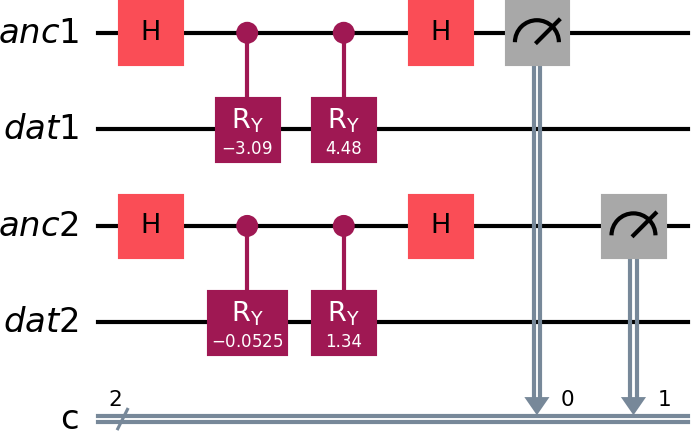}
  \end{tabular}
  \caption{Quantum circuit of the Hadamard test for cosine similarity between two-dimensional vectors by Qiskit. Example circuit for $ d = 2 $.}\label{fig2-1}
\end{figure}

For example, the quantum circuit for a two-dimensional vector is shown in Figure \ref{fig2-1}.
In this figure, we observe that the independent circuits corresponding to each vector element are arranged in parallel.

\subsection{Approximation bias}
Next, we describe the bias caused by the approximation in Eq. \ref{eq5}.
The bias is the difference between the estimated cosine similarity (Eq. \ref{eq6}) and the true value $ v \cdot w $, defined as follows:
\begin{equation}\label{eq7}
  Bias \coloneqq \left( \sum_{i=1}^{d} {\rm Re} \braket{0 | U_{i} | 0} \right) - d + 1 - \sum_{i=1}^{d} v_{i} w_{i}.
\end{equation}
Using Eq. \ref{eq7}, we obtain
\begin{equation}\label{eq8}
  \begin{split}
    Bias &= \left( \sum_{i=1}^{d} {\rm Re} \braket{0 | U_{i} | 0} \right) - d + 1 - \sum_{i=1}^{d} v_{i} w_{i}\\
    &= \left( \sum_{i=1}^{d} \sqrt{1 - v_{i}^{2}}\sqrt{1 - w_{i}^{2}} \right) - d + 1\\
    &\geq \sqrt{\sum_{i=1}^{d} 1 - v_{i}^{2}}\sqrt{\sum_{i=1}^{d} 1 - w_{i}^{2}} - d + 1\\
    &= d - 1 - d + 1\\
    &= 0\\
  \end{split}
\end{equation}
In the above, in the first inequality, we use the Cauchy-Schwarz inequality.
Therefore, $ Bias \geq 0 $, indicating a non-negative bias.

\subsection{Comparison of algorithm features}
We compare the amplitude-encoding and angle-encoding Hadamard tests in Table \ref{tab1}.
\begin{table}[htb]
  \centering
  \caption{Comparison of features for amplitude-encoding Hadamard test and angle-encoding Hadamard test. $ d $ denotes the vector size.}\label{tab1}
  \vspace{5pt}
  \begin{tabular}{ccc}\hline
    & Amplitude-encoding & Angle-encoding\\ \hline
    Circuit size (\# qubits) & $ \mathcal{O}(\log_{2}(d)) $ & $ \mathcal{O} (d) $\\
    State-preparation gate count & \# gates: $ \mathcal{O}(d) $ & \# gates: $ \mathcal{O}(d) $\\
    Circuit depth & $ \mathcal{O}({\rm poly}(d)) $ & $ \mathcal{O}(1) $\\
    Classical post-processing & No & Yes\\ \hline
  \end{tabular}
\end{table}
The main advantage of the angle-encoding Hadamard test is that its circuit depth is $ \mathcal{O}(1) $, which makes it suitable for NISQ devices.
The disadvantages are (1) approximation errors (bias) in the results and (2) the need for classical post-processing.

\section{Numerical experiments and results}\label{sec4}
To calculate the cosine similarity, two vectors of dimension $ d $ were generated by sampling each entry uniformly at random from $ [-1, 1] $, after which the vectors were normalized.
A total of 100 data samples were generated using 100 different random seeds.

\begin{figure}[htbp]
  \centering
  \begin{minipage}{10.5cm}
    \SetFigLayout{2}{2}
    \subfigure[d = 2, n = 4]{\includegraphics[width=5cm]{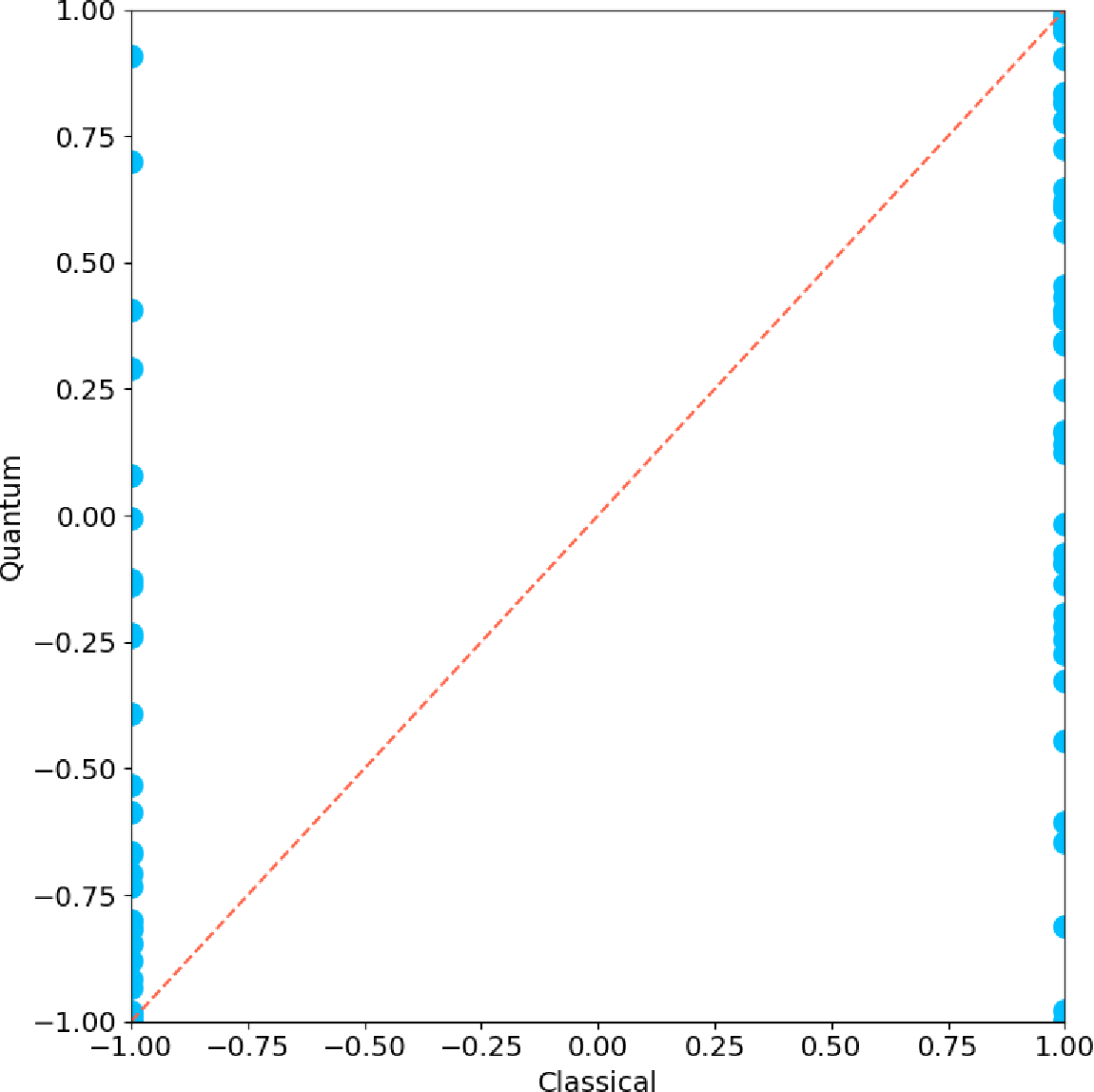}}
    \hfill
    \subfigure[d = 4, n = 8]{\includegraphics[width=5cm]{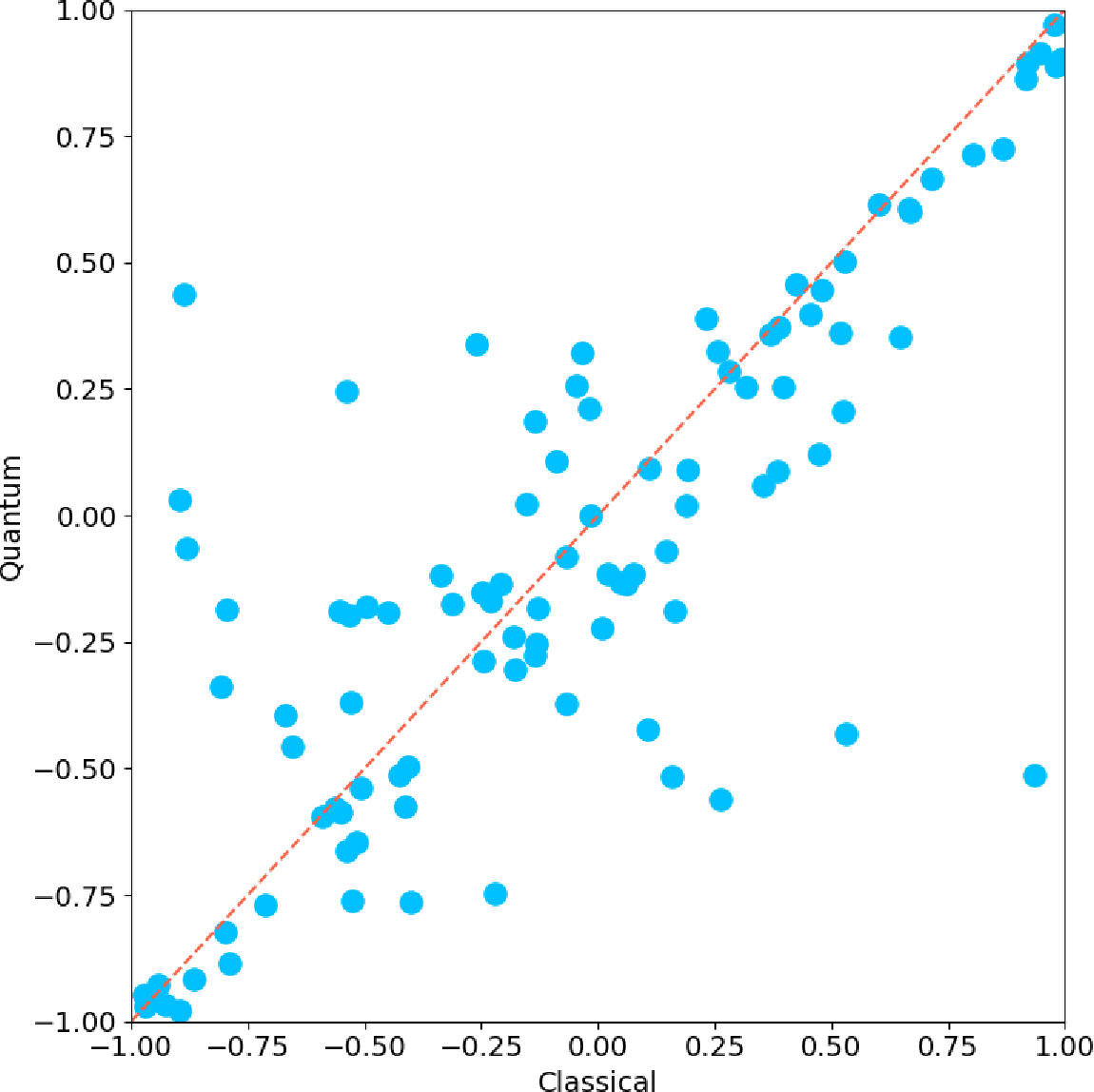}}\\
    \subfigure[d = 8, n = 16]{\includegraphics[width=5cm]{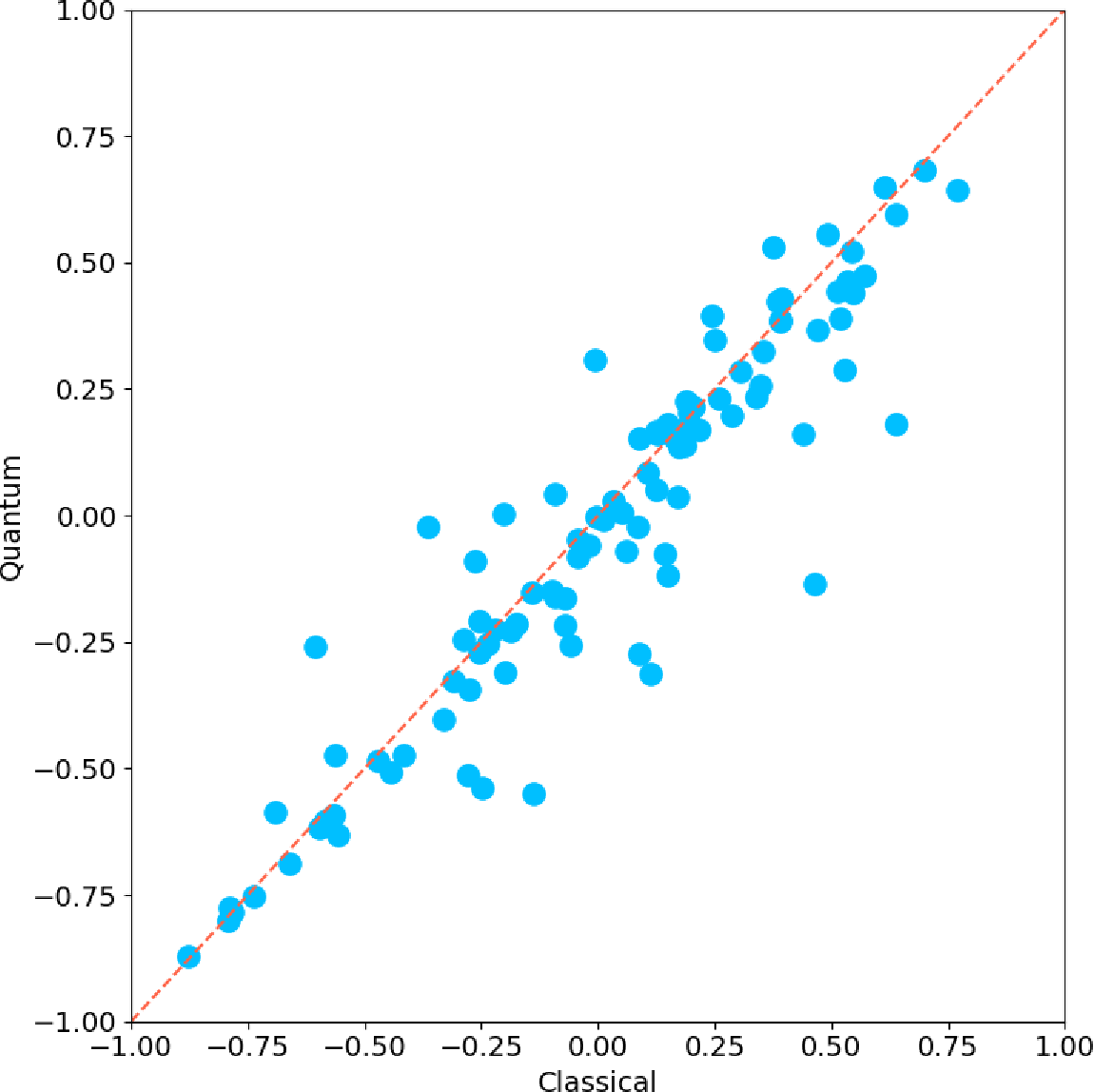}}
    \hfill
    \subfigure[d = 12, n = 24]{\includegraphics[width=5cm]{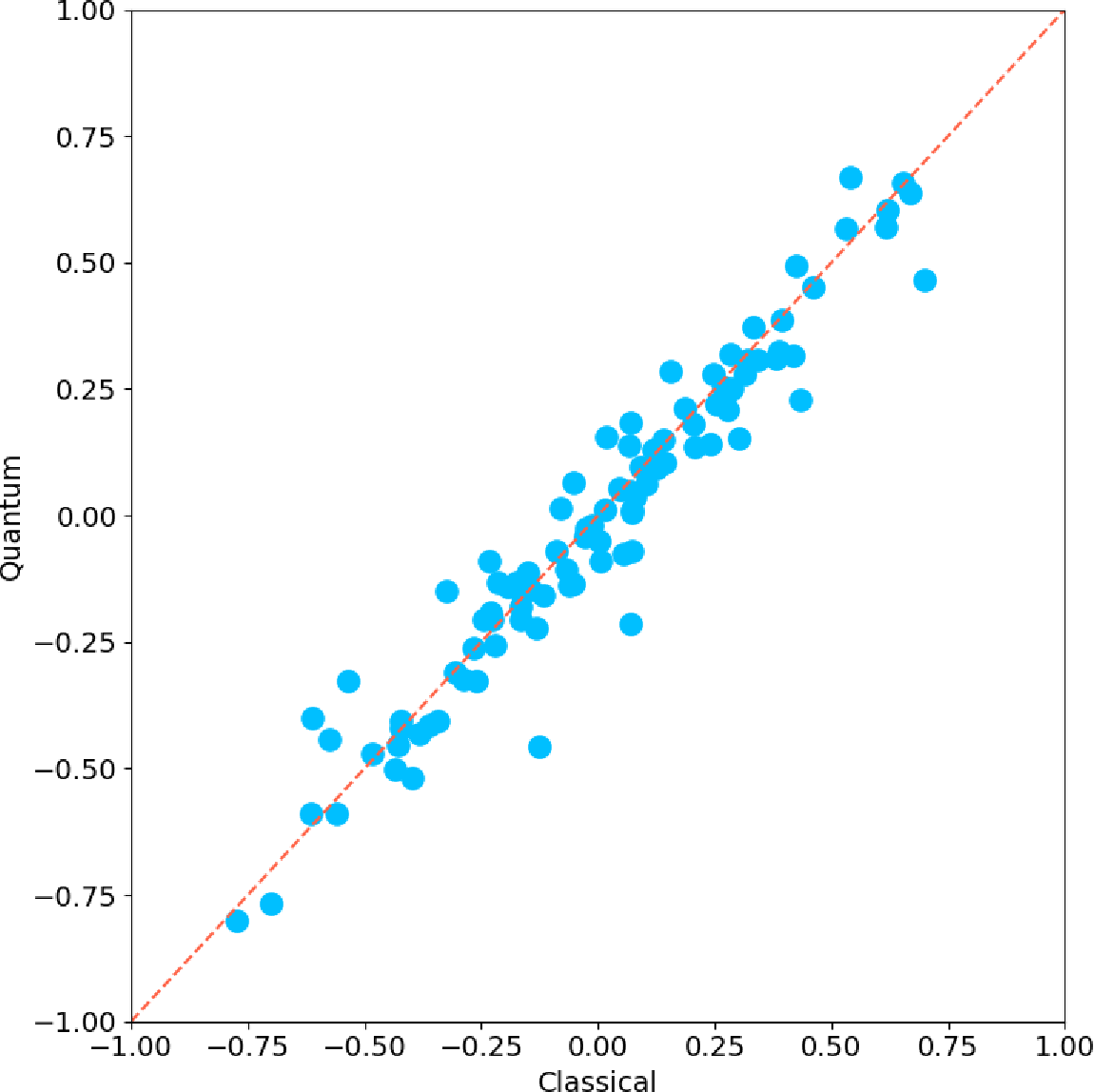}}
  \end{minipage}
  \caption{Scatter plots of cosine similarity computed classically and by the proposed algorithm}\label{fig3-1}
\end{figure}

For varying vector sizes and qubit sizes, the root mean square error (RMSE) and the correlation coefficient between the classical cosine similarity and the value estimated by the proposed algorithm were calculated.
The results are summarized in Table \ref{tab2}, where the qubit size is twice the vector size.
\begin{table}[htb]
  \centering
  \caption{RMSE and correlation coefficient for cosine similarity estimation}\label{tab2}
  \vspace{5pt}
  \begin{tabular}{cccc}\hline
    Vector size (d) & \# qubits (n) & RMSE & Correlation coefficient\\ \hline
    2 & 4 & 0.8012 & 0.6449\\
    4 & 8 & 0.3479 & 0.7857\\
    8 & 16 & 0.1500 & 0.9301\\
    12 & 24 & 0.0879 & 0.9642\\ \hline
  \end{tabular}
\end{table}
In addition, scatter plots are shown in Figure \ref{fig3-1}.
As the vector size (and thus the qubit size) increases, the RMSE decreases while the correlation coefficient increases, leading to improved scatter-plot distributions.
This is because the magnitude of each element of the vectors becomes small on average, thereby relatively reducing the approximation error in Eq. \ref{eq5}.

\begin{figure}[htbp]
  \centering
  \begin{tabular}{c}
    \includegraphics[width=8cm]{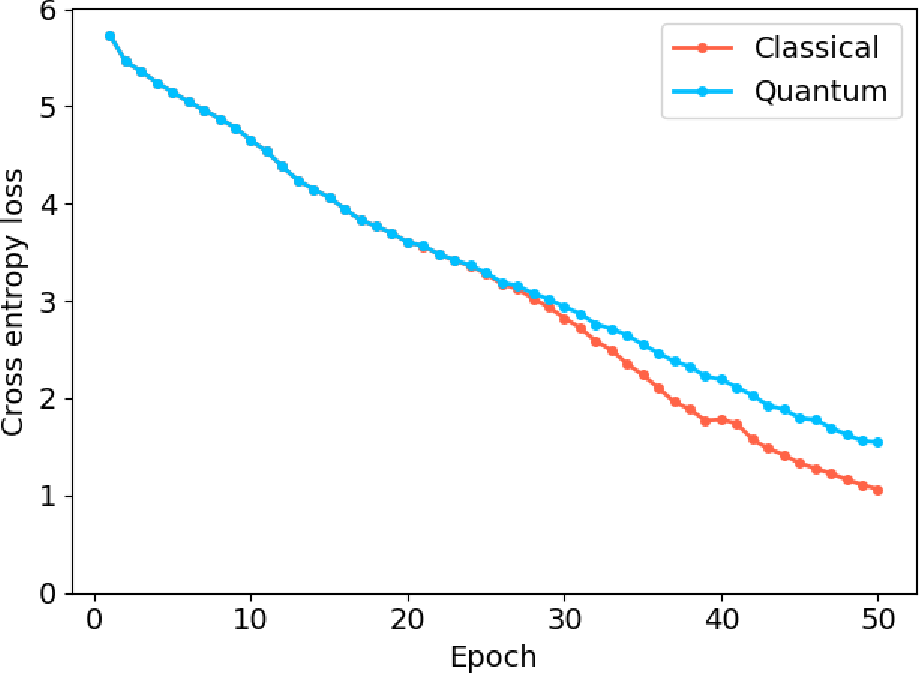}
  \end{tabular}
  \caption{Training curves of a Transformer using classical and quantum cottention mechanisms}\label{fig3-2}
\end{figure}

\clearpage
Finally, as an application, we applied the proposed algorithm to the Transformer architecture \cite{vaswani2017}, which underlies large language models.
In particular, we focused on the attention matrix in Transformers because of its relatively high computational cost, which leads to high energy consumption.
In the future, quantum computers may offer energy advantages for certain computational tasks over classical computers.
Motivated by this possibility, we propose using quantum circuits as subroutines within the Transformer.
In Transformers, scores are computed between query and key vectors, and attention weights, referred to as softmax attention, are obtained by applying the softmax function to the scaled scores.
For very long input sequences, this approach can significantly reduce memory consumption compared with softmax attention.
In \cite{mongaras2025}, the attention weights are defined using the cosine similarity between the query and key vectors.
For very long input sequences, this cosine-similarity-based attention (cottention) can significantly reduce memory consumption compared with softmax attention.
Hence, the angle-encoded Hadamard test was used to calculate cosine similarity, which we refer to as quantum cottention.
In the experiments, we used the Japanese-English portion of the JEC Basic Sentence Data.
The dataset is distributed by the Language Media Processing Lab at Kyoto University, and is licensed under CC BY 3.0 Unported.
In this study, to achieve a reasonable wall time, we set the number of training samples to 53.
The vector size in the Transformer was set to 16.
To maintain a reasonable training wall time, only eight qubits were used, and the 16-dimensional similarity computation was divided into four runs.
For the quantum circuits, we used state-vector simulations.
The training curves are shown in Figure \ref{fig3-2}.
At the early stages of training, the classical and quantum cottention mechanisms exhibit nearly identical training curves.
However, as the number of epochs increases, the discrepancy between the two cottention mechanisms gradually becomes more pronounced.
This behavior is attributable to differences in the accuracy of the cosine similarity.
As such, one possible reason why the difference becomes more apparent during training is that cosine similarity is computed at a deep stage of the Transformer's processing flow.

\section{Conclusion}\label{sec5}
In this study, we propose an angle-encoding Hadamard test to reduce the computational demand of the conventional Hadamard test.
Although the proposed algorithm yields approximate outputs compared with the conventional Hadamard test based on amplitude encoding, increasing the size of the input state (i.e., the number of qubits) leads to improved output accuracy.
Through numerical experiments, we confirmed the performance of the proposed algorithm.

\bibliographystyle{plain}
\bibliography{references}

\end{document}